\documentclass[aps,prl,twocolumn,epsf]{revtex4}

\usepackage{graphicx}
\usepackage{dcolumn}
\usepackage{bm}
\begin{document}
\title{Properties of Zero-Lag Long-Range Synchronization via Dynamical Relaying}
\author{Maria de Sousa Vieira}
\altaffiliation[Also at ]{Thomson Financial, 425 Market St., San Francisco, CA 94105. Email: mariav\_us@yahoo.com} 
\affiliation{International Center for Complex Systems, Universidade Federal do Rio Grande do Norte, CEP 59078-970, Natal, RN,  Brazil.} 
\begin{abstract}
In a recent letter, Fisher {\sl et al.} reported the phenomenon of zero-lag 
long range isochronous  synchronization via dynamical relaying 
in systems with delay [Phys. Rev. Lett. 
{\bf 97}, 123902 (2006)]. 
They reported that when one has two coupled systems A and C, 
with delay between them, then the introduction of a third element B between 
A and C will allow them to synchronize even in regions  of the parameter 
space where this was not possible without the presence of B.  
Here we study in detail the phenomenon and verify 
that in all the cases studied (including the ones reported by Fisher {\sl et al.}) 
this occurs due to the tendency of A and B and B and C to be in antiphase 
synchronization and if A is in antiphase with B and B is in antiphase with 
C, it will imply that A and C are inphase. We show this in coupled 
quadratic maps, Kuramoto and R\"ossler oscillators. We also report that there 
is a simpler configuration where the same phenomenon occurs and has nearly the  
same features of the cases studied by  Fisher {\sl et al}. 
\end{abstract}
\pacs{PACS numbers: 05.45.Xt,  05.45.Pq,  05.45.Ac.}
\maketitle
\section{Introduction}\label{s1}
Synchronization phenomenon is widely observed in nature and in artificial systems.
The first description of synchronization is believed to have been made by 
Huygens. He observed that two pendulum clocks suspended in the same wooden beam tend to synchronize in opposite swings\cite{huygens}. In more recent studies, 
synchronization has been observed in fireflies, pacemaker cells, networks of neurons, etc\cite{pikovsky}. 
In brain activity, it has been found that cognitive acts, such as face perception, is associated with 
zero-lag (i.e., isochronal) synchronization of gamma 
oscillations\cite{rodriguez}. Other studies have shown neural 
synchronization in 
widely separated areas of   
cortical regions \cite{engel,roelfsema}.

In a recent publication,  Fisher {\sl et al.}\cite{fisher} 
reported the phenomenon of 
zero-lag long range synchronization via dynamical relaying. That is, 
when one has two coupled systems, with delay between them, introducing 
a third one will allow synchronization between the two outer elements 
in regions of the parameter space where this was impossible without the 
middle element.  In their letter they showed that this occurs for 
coupled lasers (in experimental and numerical studies), coupled neurons
and stated that they observed the same phenomenon in a large variety 
of dynamical systems (such as excitable systems, oscillators and maps).   

The aim of this paper is to  provide an explanation for this phenomenon. 
We find that this occurs because two identical systems 
lose synchronization as the delay between them is increased 
in such a way that they get in antiphase synchronization. That is, 
in the case of oscillators their phase separation is $\pi $. 
When a third element is added between those elements, then 
the middle element will be in antiphase with the outer elements and 
therefore the outer elements will be inphase. 
Here we call the attention to the fact that it is not any finite delay that 
will cause loss of synchronization between two coupled systems, as we will see 
below. For loss of 
synchronization to occur the 
delay has to be in a given range of values and therefore this 
phenomenon is only important in that region. 

In their letter,  Fisher {\sl et al.} studied bidirectional coupling 
between the middle and outer elements. Here we show that nearly identical results can be 
obtained in a simpler configuration, that is, the subsystem coupling 
as studied in \cite{pecora}. We demonstrate our results in  
quadratic maps\cite{quadratic}, Kuramoto\cite{kuramoto} and coupled R\"ossler\cite{rossler} oscillators.  
In the case of Kuramoto oscillators, we show analytically that when all the oscillators are identical,  synchronization between  
the two outer elements  always occur, 
independently of the delay and coupling values.
Although the synchronization properties do not depend on if the coupling is 
bidirectional or of the subsystem type, in both cases it is necessary 
that the parameters of the outer elements be the same. 

The paper is organized as follows:
In the next section we study zero-lag synchronization via dynamical relaying in a  discrete time system, that is, the quadratic map. Then we turn our attention to this phenomenon in continuous time systems. In particular, we study the 
Kuramoto and R\"ossler oscillators. The last section is dedicated to discussion. 

\section{Quadratic Maps}
Here we study a system of coupled quadratic maps. 
For two coupled quadratic maps with delay of two time steps, we
have
\begin{eqnarray}
x_1^{t+1}&=&(1-k)f(x_1^t)  + kf(x_2^{t-2})  \nonumber \\
x_2^{t+1}&=&(1-k)f(x_2^t)  + kf(x_1^{t-2})  
\label{eq1},
\end{eqnarray}
where $f=1-ax^2$, $a \in [0,2]$, with  $f$ mapping the interval [-1,1] into itself. The case when  the coupling 
strength, $k$,  is zero has been extensively studied in the literature and 
one knows that the logistic map presents period doubling bifurcations, leading 
to a chaotic behavior as $a$ is increased\cite{quadratic}. 
In the coupling scheme 
studied in \cite{fisher} our system would be
\begin{eqnarray}
x_1^{t+1}&=&(1-k)f(x_1^t)  + kf(x_2^{t-1})  \nonumber \\
x_2^{t+1}&=&(1-k)f(x_2^t)  + k[f(x_1^{t-1}) + f(x_3^{t-1})]/2 
\label{eq2}. \\
x_3^{t+1}&=&(1-k)f(x_3^t)  + kf(x_2^{t-1}). \nonumber
\end{eqnarray}
Note that in Eq.~\ref{eq1} the delay is of two time steps and in 
Eq.~\ref{eq2} the delay is of one time step, since, as  
done by Fisher {\sl et al.}, one assumes that the delay between 
elements 1 and  2, as well as between 2 and 3 is half of the 
delay between 1 and 3. 

\begin{figure}
\centering
\includegraphics[width=3.5 in, height=4.0 in]{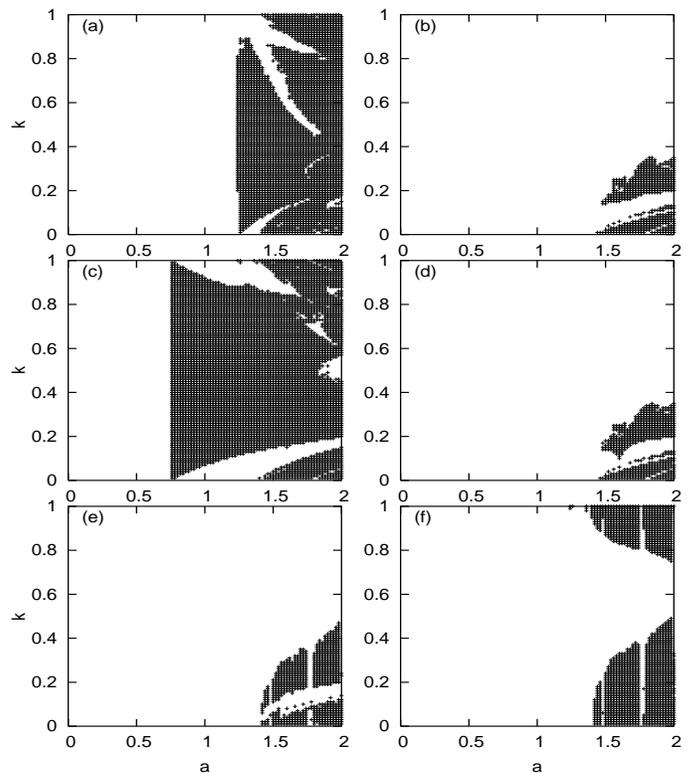}
\caption{\label{f1} 
Region of nonnegative transverse Liapunov exponent for synchronization  (shaded)  
between (a) 1 and 2 in Eq.~\ref{eq1}, (b) 1 and 3 in Eq.~\ref{eq2}, (c) 
1 and 2 in Eq.~\ref{eq2}, (d) 1 and 3 in Eq.~\ref{eq3}, 
(e) 1 and 3 for subsystem coupling without delay and 
(f) 1 and 2 for subsystem coupling without delay.  
The initial conditions used were for Eq.~\ref{eq1}: $x_1^{-2} = 0.3$, 
$x_2^{-2} = 0.1$, $x_1^{-1} = 0.25$, $x_2^{-1} = 0.05$, $x_1^{0} = 0.2$ 
and $x_2^{0}=0.0$. For Eqs.~\ref{eq2} and Eqs.~\ref{eq3} we used: $x_1^{-1} = 0.25$, 
$x_2^{-1} = 0.05$, $x_3^{-1} = 0.55$, $x_1^{0} = 0.2$, $x_2^{0} = 0.0$  
and $x_3^{0}=0.5$.  For the subsystem coupling without delay we used:
$x_1^{0} = 0.2$, $x_2^{0} = 0.0$, $x_3^{0} = 0.5$.   
A transient of 20000 iterations was discard and the Liapunov exponent 
was considered non negative if it was greater than  $10^{-4}$ 
after 50000 iterations. 
}
\end{figure}

For the system of coupled quadratic maps considered above, we noticed 
that more than one basin of attraction is present. Therefore, studying 
the region of synchronization for a given initial condition, may not 
give a full picture of the synchronization region. We observed that, as 
in the case without delay\cite{pecora}, the transverse Liapunov exponent 
correctly determines the region where synchronization is possible between 
two elements of a given system. The region of synchronization found 
in this way does not depend on the initial condition used, since 
by construction the transverse Liapunov exponent measures the separation 
(or shrinkage) of trajectories that are started nearby to each other. 
We have adapted the method of   
Benettin {\sl et al.}\cite{benettin} to study the tranverse Liapunov 
exponent of the coupled quadratic maps. If we are studying the 
synchronization between element $a$ and $b$, after the transient dies out, 
 we evolve the orbit 
of $b$ by making it slightly different from the orbit of element $a$. 
Then we verify how the orbits of $a$ and $b$ differ after one iteration. The
perturbation is renormalized in the direction of maximum growth and 
the process is repeated many times. The transverse Liapunov exponent is 
given by the average logarithm (in this paper we use base 2) of the growth 
(or shrinkage) of the perturbation along the orbit. In Figs.~\ref{f1}(a) and (b) 
we show the region where the tranverse Liapunov is non negative
(shaded area) for, respectively, elements 1 and 2 in Eq.~\ref{eq1} and 
elements 1 and 3 in Eq.~\ref{eq2}. We see that the region where synchronization 
is possible (i.e., when  the transverse Liapunov exponent is negative) is 
greatly increased with the scheme given by Eq.~\ref{eq2}. Although the 
study of synchronization between 1 and 2 in Eq.~\ref{eq2} is not the main goal 
of this work, we show in Fig.~\ref{f1}(c) for the sake of comparison, the region of  non negative 
transverse Liapunov exponent for their synchronization orbit. 
We see that synchronization 
between 1 and 2 in Eq.~\ref{eq2} occurs in a smaller region of the parameter 
space, when we compare with synchronization between 1 and 2 in Eq.~\ref{eq1} and 
between 1 and 3 in Eq.~\ref{eq2}. This shows that 1 and 3 can be synchronized even when 1 and 2 cannot. 

Our next step is to compare these results with the synchronization properties 
of 1 and 3 when the system is coupled  
via a subsystem coupling, as studied in \cite{pecora} for systems without 
delay in the context of chaotic synchronization. In this situation, the 
equations that govern the dynamics of the coupled system are 
\begin{eqnarray}
x_1^{t+1}&=&(1-k)f(x_1^t)  + kf(x_2^{t-1})  \nonumber \\
x_2^{t+1}&=&(1-k)f(x_2^t)  + kf(x_1^{t-1})  
\label{eq3} \\
x_3^{t+1}&=&(1-k)f(x_3^t)  + kf(x_2^{t-1}). \nonumber
\end{eqnarray}
We show in Fig.~\ref{f1}(d) the region of non negative transverse Liapunov exponent 
(shaded area) for the orbits of elements 1 and 3. When compared with Fig.~\ref{f1}(b) one sees that the two shaded regions are nearly identical to each other, 
showing that there is basically no difference of the parameter region 
where synchronization can occur between 1 and 3 in one coupling versus 
the other. We have noticed only very minor differences 
in the parameter region of 
synchronization between 1 and 2 in Eq.~\ref{eq3} 
when compared with Fig.~\ref{f1}(c).  
For the sake of completeness, we also show in Figs.~\ref{f1}(e) 
and (f)   
the region of non negative Liapunov exponent between 1 and 3 and 1 and 2, 
respectively, for a system without delay (for subsystem coupling). 
When compared with the case with delay, 
we see that delay helps in the synchronization of 1 and 3 (since the 
transverse Liapunov is negative in a wider area) and it is detrimental 
for synchronization between 1 and 2.

\begin{figure}
\centering
\includegraphics[width=3.5 in, height=2.0 in]{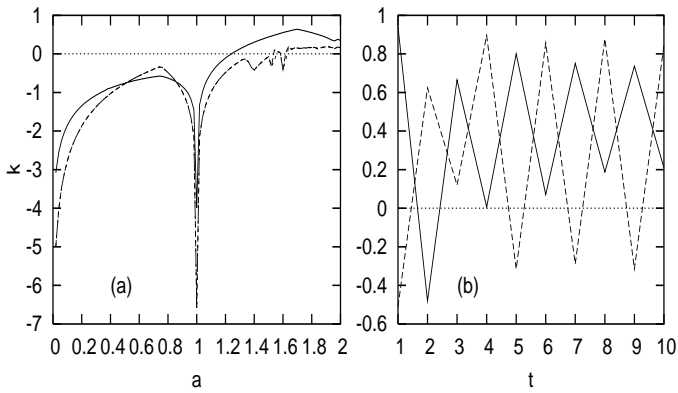}
\caption{\label{f2} 
(a) The transverse Liapunov exponent for the synchronization orbit of 1 and 
2 (solid) in  the coupled quadratic maps for the subsystem coupling.  
The dashed lines are the transverse Liapunov exponent between 1 and 3 for 
the subsystem and bidirectional coupling. (b) Time evolution 
of $x_1$ (solid) and $x_2$ (dashed) for parameter values where 
synchronization between 1 and 3 occurs. 
The parameter values are in (a) $k=0.2$ and in (b) $k=0.2$, 
$a = 1.4$. The initial conditions are the same as in Fig.~\ref{f1}.   
}
\end{figure}

We have studied the magnitude of the transverse 
Liapunov to find out which coupling, bidirectional or subsystem,  
results in faster synchronization. The more negative the 
transverse Liapunov is, the faster is the synchronization. 
As shown in Fig.~\ref{f2}(a) (dashed lines)  the Liapunov
exponent is essentially the same independently of the coupling type.  
In Fig.~\ref{f2}(b) we show the temporal evolution (for the subsystem coupling) 
of the orbit  of element 1 and the orbit of element 2 in the region where 
1 synchronizes with 3. We see that they are in antiphase synchronization, i.e., moving in opposite directions. 
Thus, 1 is in antiphase with 2, which in turn is in antiphase with 3 and 
consequently 1 is inphase with 3. So, this is the mechanism by which 
the relaying element, 2,   helps in the synchronization our the 
outer elements, 1 and 3.  

\section{Kuramoto Oscillators}

As in the case of discrete time system, we have found that the synchronization properties of 
the subsystem coupling are nearly identical to the synchronization properties 
of the bidirectional coupling.  Because the subsystem configuration is 
simpler and allow us to perform some analytical studies in the Kuramoto 
system we will concentrate from now on mostly on the subsystem case. 

\begin{figure}
\centering
\includegraphics[width=3.5 in, height=3.0 in]{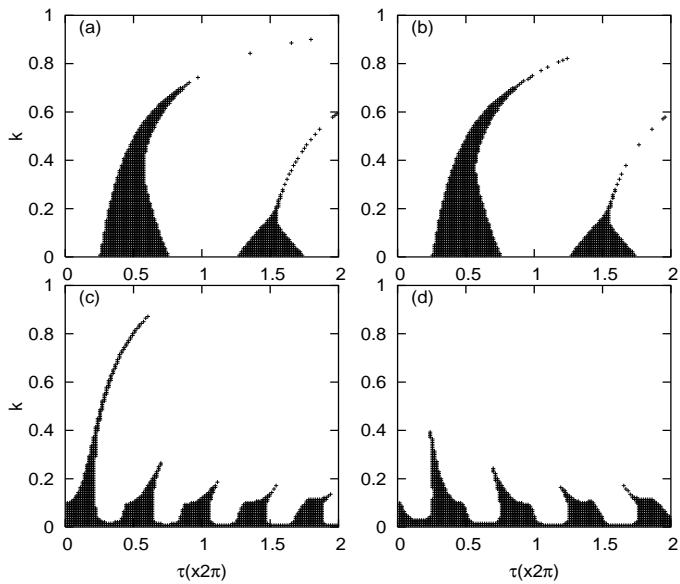}
\caption{\label{f3} 
Region where isochronal synchronization  
occurs (white) between oscillator 1 and 2 for the Kuramoto oscillators  
and when their phase separation 
is $\pi $ (shaded) for (a) subsystem 
coupling and (b) bidirectional coupling, 
with $\omega = \omega ' = 1$. Region where isochronal synchronization  
occurs (white) between oscillator 1 and 3 
for the subsystem coupling with (c)    
$\omega =1.2$, $ \omega ' = 1$, (d) $\omega =1$, $ \omega ' = 1.2$. 
The initial conditions used were
$\phi_1=0.1$, $\phi_2= 0.12$, $\phi_3=0.14$ 
and $\phi_{1,2,3}(-\tau) =\phi_{1,2,3}(t=0) + 0.001\tau$. 
}
\end{figure}

The equations for the subsystem coupling in coupled Kuramoto oscillators 
are given by 
\begin{eqnarray}
\dot \phi_1&=& \omega - k\sin(\phi_1 - \phi_2^{\tau})  \nonumber \\
\dot \phi_2&=& \omega' - k\sin(\phi_2 - \phi_1^{\tau})  
\label{eq4} \\
\dot \phi_3&=& \omega   - k\sin(\phi_3 - \phi_2^{\tau}), \nonumber
\end{eqnarray}
where we have used the notation $\phi^{\tau}$ to denote $\phi(t-\tau)$
and $\phi$ to denote  $\phi (t)$.  
We see that oscillators 1 and 2 are uncoupled from oscillator 3 and 
in this way the analysis of two coupled oscillators is helpful in the 
understanding of the system as a whole.  
A system of two coupled Kuramoto oscillators with delay was studied by 
Schuster and Wagner\cite{schuster} and is governed by  
\begin{eqnarray}
\dot \phi_1&=& \omega   - k\sin(\phi_1 - \phi_2^{\tau})  \nonumber \\
\dot \phi_2&=& \omega' - k\sin(\phi_2 - \phi_1^{\tau}) 
\label{eq5} 
\end{eqnarray}
For the sake of completeness, here we reproduce the main results of 
\cite{schuster}. They assumed that the synchronizing solution for 
the two coupled oscillators is given by 
\begin{equation}
\phi_{1,2} = \psi \pm \alpha/2,
\label{eq6}
\end{equation}
where $\psi$ is the common dependent phase and $\alpha $ is a constant
phase shift. Inserting this into Eq.~\ref{eq5} one gets  
\begin{equation}
\cos(\psi - \psi^{\tau}) = (\omega - \omega')/(2k\sin(\alpha)) = const,
\label{eq7}
\end{equation}
which is fulfilled if $\psi = \Omega t$. Thus, the synchronized solution 
is given by $\phi_{1,2}= \Omega t \pm \alpha/2$. Inserting this solution 
into Eq.~\ref{eq5} we get that the synchronizing frequency is given by 
the zeros of  
\begin{equation}
f(\Omega) = \overline \omega - \Omega - k \tan(\Omega \tau)\sqrt {\cos ^2(\Omega \tau) - k_c^2/k^2},  
\label{eq8}
\end{equation}
where $k_c \equiv (\omega - \omega')/2$, $\overline \omega \equiv (\omega_1 + \omega_2)/2$.  
It is found that multiple solutions are possible for $\Omega $ when delay 
is present\cite{schuster}. This is in contrast with the case without delay. 
The phase shift that belongs to the synchronized solution is  
\begin{eqnarray}
\alpha &=& arcsin(k_c/k\cos(\Omega \tau)), \ \ \ {\rm if} \ \ \cos(\Omega \tau)>0, \nonumber \\
\alpha &=& \pi - arcsin(k_c/k\cos(\Omega \tau)), \ \ \ {\rm otherwise}.   
\label{eq9} 
\end{eqnarray}
If $\omega = \omega '$ the above equations will reduce  to 
\begin{equation}
f(\Omega) = \omega - \Omega - k \sin(\Omega \tau),
\label{eq10}
\end{equation}
and 
\begin{eqnarray}
\alpha &=& 0, \ \ \ {\rm if} \ \ \cos(\Omega \tau)>0, \nonumber \\
\alpha &=& \pi. \ \ \ {\rm otherwise}.
\label{eq11} 
\end{eqnarray}
In Fig.~\ref{f3}(a)  we show the region where synchronization between 
oscillator 1 and 2 occurs (white) and when it does not occur for a system 
with $\omega = \omega ' = 1$. The phase separation in this case is $\pi $ as shown 
above.  
The region of non synchronization is in the shape of "tongues" with 
its center being around  $n\pi/2 $, where $n$ is an odd number.  

We have found that when $\omega = \omega '$, oscillator 1 and 3 {\sl always} synchronize, independently of the delay and coupling values.  We can show this in the following way:   
give a small perturbation $\Delta $, where $\Delta $ is positively 
defined,  to the orbit of  
oscillator 1, so that the initial condition for 
oscillator 3 is given by $\phi_3 = \phi_1 + \Delta $.  
Now 
subtracting 
the first from the  third lines of Eq.~\ref{eq4} we have 
\begin{equation}
\dot \phi_3 - \dot \phi_1 = k[\sin(\phi_2^{\tau} - \phi_3) - \sin(\phi_2^{\tau} - \phi_1)],  
\label{eq10}
\end{equation}
Using  $\Delta $ as defined above and the Taylor series expansion we get from Eq.~\ref{eq7} after a 
straight forward algebra,  
\begin{equation}
\dot \Delta = -k\Delta\cos(\phi_2^{\tau} - \phi_1)= -k\Delta\cos(\Omega \tau - \alpha),  
\label{eq11}
\end{equation}
We have seen that if $\cos(\Omega \tau) > 0$, then $\alpha = 0$ and if 
$\cos (\Omega \tau \le  0)$, then $\alpha = \pi$. Both cases result in  
$\dot \Delta /\Delta = -k |\cos (\Omega \tau )| < 0$.
Consequently, the separation $\Delta $ between the orbits of oscillator 1 and 3 
decreases as time evolves, implying that they tend to synchronize for any 
value of the parameter space (with $k>0$). We also notice that there is 
region of the parameter space where the phase differences 
(say, $0$ or $\pi$ in the case 
of identical oscillators) do not change if the delay is varied in that region. 
Consequently, we expect that even if the delay between 1 and 2 is not exactly the same as between 2 and 3, zero-lag long range isochronal synchronization between 1 and 3  would still be possible.  

For the sake of comparison, we show in Fig.~\ref{f3}(b) the region of synchronization between 1 and 2 (white) for the bidirectional coupling, that is, 
\begin{eqnarray}
\dot \phi_1&=& \omega   - k\sin(\phi_1 - \phi_2^{\tau})  \nonumber \\
\dot \phi_2&=& \omega' - k[\sin(\phi_2 - \phi_1^{\tau}) +\sin(\phi_2 - \phi_3^{\tau}) ]/2 \\
\label{eq12} 
\dot \phi_3&=& \omega   - k\sin(\phi_3 - \phi_2^{\tau}). \nonumber
\end{eqnarray}
We see that Figs.~\ref{f3} (a) and (b) are almost identical, confirming 
once more that there is basically no  distinction between the synchronization 
properties of the subsystem and bidirectional couplings. 

We have also considered the case in which $\omega \ne \omega '$. This is 
displayed in Figs.~\ref{f3} (c) and (d). When $\omega \ne \omega '$ there is 
no isochronal synchronization between 1 and 2, since there is a constant 
phase difference between the oscillators\cite{schuster}. However, synchronization between 1 and 3 is possible in a large area of the parameter space. 
The direction  
of the "tongues" where synchronization does not occur depend on 
having $\omega > \omega ' $ or the other way around. 
However, the area of non synchronization varies little 
from one case to the other.   

We have looked for even simpler configurations  
where the zero-lag long range isochronal synchronization would occur, such 
as removing one of the links between element 1 and 2 in the subsystem coupling, but we found that such a kind  synchronization is not possible in those 
simpler configurations.  

\section{R\"ossler  Oscillators}

We have chosen R\"ossler chaotic oscillators to show  
that also in this chaotic system,  governed by continuous time equations,  the  
phenomenon reported in \cite{fisher} is  
due to the fact that inphase synchronization coupled with another inphase synchronization results in isochronal 
synchronization.  The equation of R\"ossler oscillator with subsystem 
coupling are given by  
\begin{eqnarray}
\dot x_i&=& -\omega y_i - z_i - k(x_i-x_j^{\tau})  \nonumber \\
\dot y_i&=& \omega x_i + ay_i -k(y_i-y_j^{\tau})
\label{eq13}\\ 
\dot z_i&=& b +z(x_i-c)-k(z_i-z_j^{\tau}), \nonumber
\end{eqnarray}
where  $i=1,2,3$, and when 
$i=1,j=2$, $i=2,j=1$ and $i=3,j=2$.  In this paper we have used, 
$\omega = 1$, $a=0.15$, $b=0.4$ and $c=8.5$.    

\begin{figure}
\includegraphics[width=3.5 in, height=4.0 in]{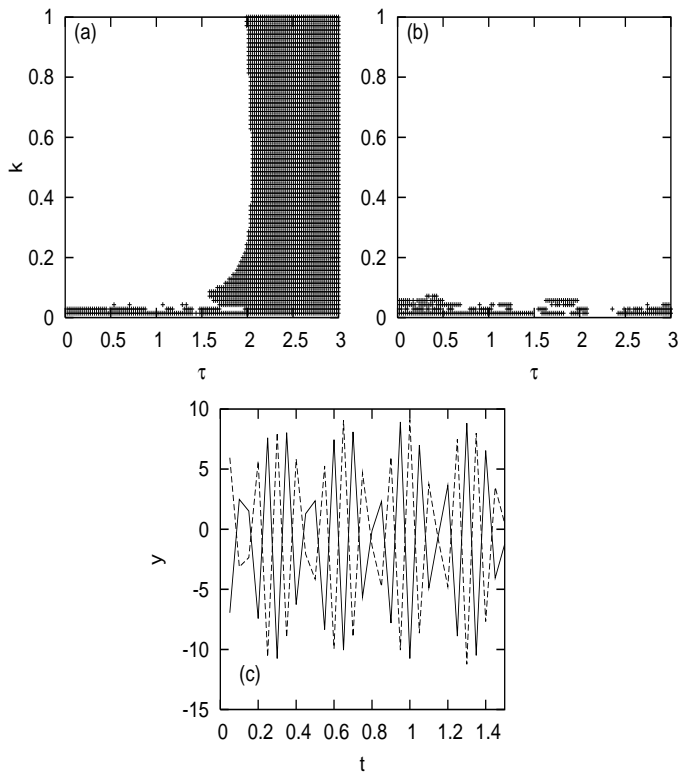}
\caption{\label{f4} 
Region where synchronization occurs (white) and when it does not (shaded) 
between 
(a) 1 and 2 and (b) 1 and 3 for the R\"ossler coupled oscillators with subsystem coupling.  
(c) Temporal evolution of $y_1(t)$ (solid) and $y_2(t)$ (dashed) when $\tau =2.5$ and $k=0.2$.
The parameter values were $\omega = 1$, $a = 0.15$, $b = 0.4 $ and $c=8.5$, 
and   
the initial conditions where 
$x_1=0.1$, $y_1= 0.12$, $z_1=0.14$, $x_2=0.1$, $y_2= 0.22$, $z_2=0.24$, $x_3=0.1$, $y_3= 0.12$, 
$z_1=0.14$ and $x,y,z(-\tau) = x,y,z(t=0) + 0.001\tau$. 
(c) Temporal evolution of $x_1(t)$ (solid) and $x_2(t)$ (dashed) when $\tau =2.5$ and $k=0.2$.
}
\end{figure}

It is shown in Figs.~\ref{f4}(a) and (b)  the region 
where synchronization happens (white) 
between oscillators 1 and 2 and between oscillators 1 and 3, respectively. 
We see once more that the presence of a middle element increases the 
region where synchronization is possible.   
In Fig.~\ref{f4}(c) we show the time evolution of $y_1(t)$ (solid) and 
$y_2(t)$ (dashed) for a region where oscillator 1 and 3 synchronize, that is, 
$\tau = 2.5$ and $k=0.2$. Once 
again we see the antiphase synchronization observed  
in the previous sections.

\section{Discussion}

By studying the results presented in the paper 
by Fisher {\sl et al.},  
we see the origin of for zero-lag long range isochronal synchronization 
they observed 
is the same we 
have found. 
If we take a look on Figs. 2(E), 2(F), 3(E) and 3(F) of \cite{fisher}  
we see that they shifted the time series of the middle element  by 
the delay value in order 
to facilitate the comparison with the time series of the outer elements. 
When they did this shift, the time series all the elements nearly matched each other.
But it turns out that,  apparently not realized by them, the  
value  $\tau_c$ is the result of this antiphase mechanism mentioned in 
the previous paragraphs. In our paper we have clarified this point, and 
showed that there is a simpler configuration (the subsystem coupling) where 
nearly identical properties are observed when compared with the bidirectional coupling. We have also been able to perform analytical studies and show that 
for the Kuramoto system with identical oscillators, 
synchronization between the outer elements always occurs, independently of the parameter values.

\end{document}